\documentclass[preprint2]{aastex}

\newcommand{\be}{\begin{equation}}
\newcommand{\ee}{\end{equation}}

\newcommand{\aox}{$\alpha_{\rm ox}$}

\newcommand{\plm}{$\pm$}

\newcommand{\swift}{{\it Swift}}
\newcommand{\xmm}{{\it XMM-Newton}}

\shorttitle{Swift observation of Mkn 335}
\shortauthors{Grupe et al.}

\begin{document}


\def\etal{{\it et\thinspace al.}\ }
\def\alp{{$\alpha$}\ }
\def\al2{{$\alpha^2$}\ }

%
%
%


\title{Discovery of the Narrow-Line Seyfert 1 galaxy Mkn 335 in an historical low X-ray flux state 
}


\author{Dirk Grupe\altaffilmark{1}
\email{grupe@astro.psu.edu},
Stefanie Komossa\altaffilmark{2},
Luigi C. Gallo\altaffilmark{3}
}

\altaffiltext{1}{Department of Astronomy and Astrophysics, Pennsylvania State
University, 525 Davey Lab, University Park, PA 16802} 

\altaffiltext{2}{Max-Planck-Institut f\"ur extraterrestrische Physik, Giessenbachstr., D-85748 Garching,
Germany; email: skomossa@mpe.mpg.de}

\altaffiltext{3}{SUPA, School of Physics and Astronomy, 
University of St. Andrews, North Haugh, St. Andrews, Fife KY16 9SS;
email: lgallo:ap.stmarys.ca}




\begin{abstract}
We report the discovery of the Narrow-Line Seyfert 1 galaxy Mkn 335
in an extremely low X-ray state.
A comparison of 
\swift\ observations obtained in May and June/July 2007 with all previous X-ray
observations between 1971 to 2006
show the AGN to have diminished in flux by a factor of more
than 30, the lowest
X-ray flux Mkn 335 has ever been observed in.
The \swift\
observations show an extremely hard X-ray spectrum
at energies above 2 keV. 
 Possible interpretations include
partial covering absorption or X-ray reflection from the disk.
In this letter we consider the partial covering 
interpretation.
The \swift\ observations can be well fit
by a strong
partial covering absorber with varying absorption
column density
($N_{\rm H}= 1-4\times
10^{23}$ cm$^{-2}$) and a covering fraction $f_c$=0.9 - 1.
When corrected for intrinsic absorption, the X-ray flux
of Mkn 335 varies by only
factors of 4-6.
In the UV Mkn 335 shows 
variability in the order of 0.2 mag.  
We discuss the similarity of Mkn 335 with the highly variable NLS1 WPVS007,
and speculate about a possible link between NLS1 galaxies and 
broad-absorption line quasars.
\end{abstract}


\section{Introduction}

Since the mid 1980s Narrow-Line Seyfert 1 galaxies \citep[NLS1s; ][]{oster85}
have become a field of extensive study in
AGN science. NLS1s are crucial for our understanding of the AGN phenomenon,
because they are most likely AGN at an early stage \citep[e.g. ][]{grupe04b}.
 They possess relatively low-mass black holes and high Eddington ratios
 $L/L_{\rm Edd}$. 
NLS1s are characterized by extreme properties, such as steep 
soft and hard X-ray spectra, strong X-ray variability, and
strong optical Fe II emission 
\citep[e.g. ][]{bol96, lei99a, lei99b, grupe01, grupe04, bor92}.

The NLS1 Mkn 335
($\alpha_{2000}$ = $00^{\rm h} 06^{\rm m} 19.^{\rm s}5$, 
$\delta_{2000}$ = $+20^{\circ} 12' 11 \farcs 0$, z=0.026) is a well-known bright
soft X-ray AGN and has been the target of most X-ray observatories. 
It was seen as a bright X-ray AGN by UHURU \citep{tananbaum78} and EINSTEIN \citep{halpern82}.
\citet{pounds87} reported a strong 
soft X-ray excess found in the EXOSAT spectrum, which was confirmed by BBXRT observations
\citep{turner93}.
GINGA observations of Mkn 335 suggested the presence of a warm absorber in the
 source \citep{nandra94}. 
During ROSAT observations it also appeared bright and with a strong soft 
X-ray excess \citep{grupe01}. 
The X-ray spectrum during the 1993 ASCA observation \citep{george00}
was either interpreted by the presence of a warm absorber \citep{lei99b} or
by X-ray reflection on the disk \citep{ballantyne01}. Beppo-SAX observations of Mkn 335 
also confirm the presence of a strong soft
X-ray excess \citep{bianchi01} and a small or moderate Compton reflection component.
\xmm\ observed Mkn 335 in 2000 and again in 2006
\citep{gondoin02,longinotti07a,longinotti07b,oneill07}.
 Mkn 335 is exceptional in showing evidence for
  an unusually broad wing in the iron line \citep{longinotti07a}.
 The wing is required, if the XMM spectrum
  is explained in terms of reflection; it is not, if
  a partial covering interpretation is adopted. 
  High-amplitude variability provides
  important new constraints to distinguish between
  these different spectral models. 
Mkn 335 was observed by \swift\ \citep{gehrels04}  in 2007 May and appeared to
be dramatically fainter in X-rays than seen in all previous observations.
In this letter we report on this historical low X-ray flux state of Mkn 335 and
compared the continuum properties of the \swift\ 
with previous \xmm\ observations.

Throughout the paper spectral indexes are denoted as energy spectral indexes
with
$F_{\nu} \propto \nu^{-\alpha}$. Luminosities are calculated assuming a $\Lambda$CDM
cosmology with $\Omega_{\rm M}$=0.27, $\Omega_{\Lambda}$=0.73 and a Hubble
constant of $H_0$=75 km s$^{-1}$ Mpc$^{-1}$ corresponding to  a luminosity distance D=105 Mpc.
All errors are 90\% confidence unless stated otherwise.

\section{\label{observe} Observations and data reduction}

\swift\ observed Mkn 335 on 2007 May 17 and 25 and June 30 to July 02 
for 4.8, 8.2, and 8.7 ks (Table\,\ref{obs_log}), respectively, with its 
X-Ray Telescope (XRT) in Photon Counting mode (PC mode)
and in all 6 filters of the UV-Optical Telescope (UVOT). 
X-ray data were reduced with the task {\it xrtpipeline} version 0.11.4.
Source and background photons were extracted
with {\it XSELECT} version 2.4, from
circles with radii of 47$^{''}$ and 189$^{''}$, respectively. 
 The spectral data were re-binned with at least 20 photons per bin
{\it grppha} version 3.0.0. 
The 0.3-10.0 keV spectra were 
analyzed with {\it XSPEC} version 12.3.1x \citep{arnaud96}.   
The auxiliary response files were created with {\it
xrtmkarf} and corrected using the exposure maps,
and the standard response matrix {\it swxpc0to12\_20010101v008.rmf}.

The UVOT data were coadded for each segment in each filter with the UVOT 
task {\it uvotimsum} version 1.3. 
Source photons in all filters
 were selected in a circle with a radius of 5$^{''}$. 
 UVOT magnitudes and fluxes were measured with the task {\it uvotsource} version
 3. 
The UVOT data were corrected for Galactic reddening \citep[$E_{\rm B-V}=0.035$; ][]{sfd98}.

\xmm\ observed Mkn 335 in 2000 and 2006 for 37ks and 133 ks, respectively (see Table\,\ref{obs_log}). 
During the 2000 observation the Optical Monitor (OM) did
photometry in the V, B, U, and M2 filters. 
During the 2006 observation the UV grism was used exclusively. 
The \xmm\ EPIC pn data were analyzed using the 
XMMSAS version {\it xmmsas\_20060628\_1801-7.0.0}. The 2000 amd 2006 observations were performed in full-frame and small 
window, respectively. Because the 2000 observation was severely affected by pileup, photons 
from a 20$^{''}$ source-centered circle were excluded. 
The source photons in the 2006 pn data were selected in a radius of 1$^{'}$ and
background photons of both observations from a source-free region close by with the same radius. 
The spectra were rebinned with 100 photons per bin. 
In order to compare the photometry in the OM with the UVOT we selected 5 field stars with 
similar brightness in V as Mkn 335. Only in B and M2
the OM magnitudes had to be adjusted  by $-$0.10
mag and $+$0.30 mag, respectively.

\section{\label{results} Results}

\begin{figure}
\epsscale{0.8}
\plotone{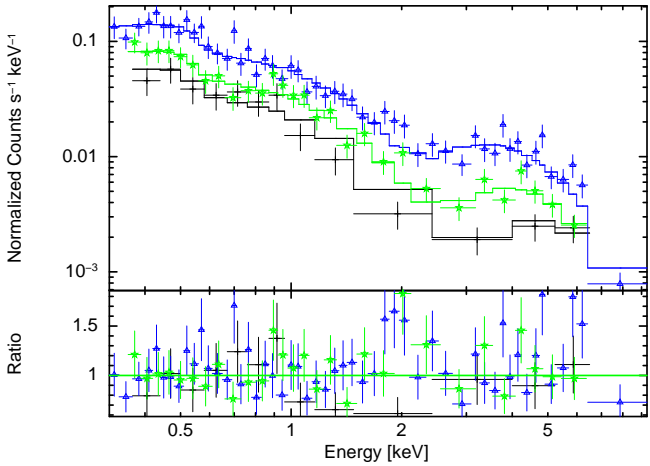}
\caption{\label{mkn335_xspec} \swift-XRT spectra of Mkn 335 fitted with a power
law model with partial covering absorber as listed in Table\,\ref{xray_res}.
The black spectrum displays the \swift\ segment 001 
spectrum, the segment 002 spectrum is in blue (triangles), 
and the segment 003-005 spectrum in green (stars). 
}
\end{figure}

None of the \xmm\ and \swift\  spectra can be fitted by a single absorbed
power law model. 
 In the literature a variety of spectral models have been
applied to the X-ray data of Mkn 335, including warm absorption \citep{lei99a, nandra94}, partial covering
\citep{tanaka05} and reflection 
\citep{gondoin02,ballantyne01, crummy06, longinotti07a,longinotti07b}. 
Fits with a warm absorber model ({\it absori}) and blackbody plus power law model
yield unacceptable results. 
While fits to the \xmm\ 2000 data yield acceptable fits by using an absorbed broken power law model with the 
absorption
column density fixed to the Galactic value \citep[3.96$\times 10^{20}$ cm$^{-2}$; ][]{dic90}, 
the \xmm\ 2006 and \swift\
spectra require additional components. We used a partial covering absorber model with an underlying
power law and
broken power law spectral models. 
Table\,\ref{xray_res} summarizes the results from the X-ray spectral analysis.
Figure\,\ref{mkn335_xspec} displays the \swift\ spectra fitted with a power law and partial covering
absorber. 
 Fits to each spectrum were first performed separately.  Subsequently all the \swift\ spectra were fitted
 simultaneously in XSPEC  
 with the power law spectral slopes
tied and the absorber parameters and the normalizations left to vary.
The results are listed in Table\,\ref{xray_res} and suggest
a development of the partial covering absorber over time. The most dramatic change is from the 2006 \xmm\ to the first \swift\
observation when the absorber became nearly opaque and  only 2\% of the X-ray emission can be seen directly.
In this case the absorption column density changes from $5\times 10^{23}$ cm$^{-2}$ with a covering fraction of 0.45 
during the 2006 \xmm\ observation to about 
 $4\times 10^{23}$ cm$^{-2}$ and a covering fraction of 0.98 during the first Swift observation. 

 We fitted all three \swift\ spectra simultaneously in XSPEC by tying the covering fraction $f_c$ 
 and spectral indices together. This fit suggests a
 change in the absorption column density $N_{\rm H,pcf}$ of the partial covering absorber by a factor of 2 within a week between the 2007 May 17 and 25 observations.
Alternatively, we also fitted the spectra with $N_{\rm H,pcf}$ tied and $f_c$ left as a free parameter. An F-test gives an F-value of 7.8 that these two
fits are different and a probability P=0.006 of a random result. Leaving $N_{\rm H,pcf}$ free gives a significantly better result than leaving $f_c$
free to vary.
In the rest-frame 0.2-2.0 keV band the observed fluxes (only corrected for Galactic absorption) seem to be highly variable and
between the \xmm\ 2000 and the first \swift\ observation we found variability by a factor of 30. However, when correcting also for
intrinsic absorption the unabsorbed restframe 0.2-2.0 keV fluxes 
from ROSAT to Swift
are comparable. During the ROSAT All-Sky Survey observation a flux of 4$\times 10^{-14}$ W m$^{-2}$ was found \citep{grupe01}.
Correcting for a partial covering absorber in the  \xmm\ and \swift\ spectra we found that the flux varied only by factors of 4-6
as listed in Table\,\ref{xray_res}.

The X-ray spectra of Mkn 335 in the higher and more typical flux state can
be well described as arising from an incident power law and reflection
component \citep[e.g. ][]{crummy06, longinotti07a}.
 However, the
low-flux spectra are difficult to reproduce by simply rescaling the high-state
models or by varying the relative contribution of each component.  A modified
reflection model, which self-consistently describes the high- and low-flux
states is being investigated and is presented in Gallo et al. (in prep).

\begin{figure}
\epsscale{0.75}
\plotone{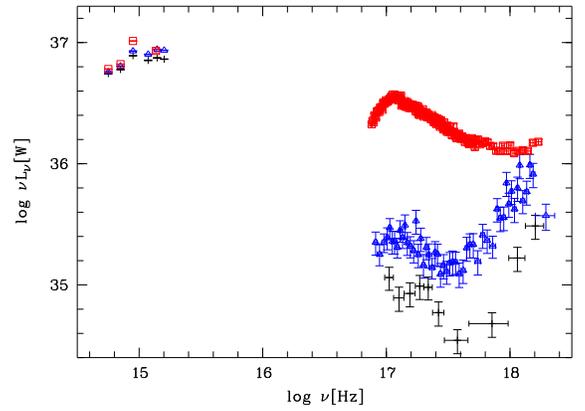}
\caption{\label{mkn335_sed}  Spectral energy distributions of Mkn 335. 
The black crosses are from the \swift observation segment 001, 
blue triangles form
segment 002, and red squares from the 2000 \xmm\ observation. The \swift\ segments 003-005
spectrum would be between the 001 and 002 spectra.
}
\end{figure}

As shown by the spectral energy distribution (SED)
in Figure\,\ref{mkn335_sed} there was no dramatic variability in the UV data
between the 2000 \xmm\ OM and 2007 \swift\ UVOT observations, although during the 2007 May 17th 
observation Mkn 335 was about 0.2 mag fainter. 
The UV/optical 
spectral slopes are on the order of
$\alpha_{\rm UV}=-0.4$, except for the 2007 May 17 observation when it was $\alpha_{\rm UV}=-0.3$.
The UV to X-ray spectral slope \aox\footnote{The
X-ray loudness is defined by \citet{tananbaum79} as \aox=--0.384
log($f_{\rm 2keV}/f_{2500\AA}$).} 
was significantly steeper during the \swift\ observations with \aox=1.91 and 1.65 during the \swift\ segments 001 
and  002, respectively. During the 2000 \xmm\ observation, however, 
an \aox=1.32 was measured, consistent with the value given by \citet{gallo06}.

\section{\label{discuss} Discussion}

We reported the \swift\ observations of the NLS1 Mkn 335 when it
 was in its lowest X-ray flux state
ever observed. 
Historically, Mkn 335 has exhibited X-ray variability by about a factor
of a few \citep[e.g. ][]{turner93, markowitz04},
although the source has always\footnote{Except for an episode in 1983 when it
had a rather low X-ray flux during its EXOSAT observation as reported by
\citet{pounds87}}
remained rather bright 
at least until the last X-ray observation with
{\it Suzaku} in 2006 June (J. Larsson, 2007 priv. comm.).
However, sometime between 2006 June and 2007
May the observed flux dropped
by a factor of more than 30 including a dramatic change in its SED.
The X-ray spectrum has become progressively more complex 
as the X-ray flux has diminished, indicative of either absorption or reflection
\citep[e.g. ][]{gallo06}.
The 2-10 keV high-flux spectrum in
2000 did  not appear overly complex and the high-energy continuum could be simply fitted with a
power law. The
2006 \xmm\ data, however, can be fitted with
a partial covering absorber model \citep[see also ][]{oneill07},
 suggesting that the absorber started moving in the 
line of sight before 2006
January. 

Partial covering of the central light source has been invoked
since the early days of AGN X-ray spectroscopy \citep[e.g. ][]{holt80}, 
and quite often to describe the X-ray spectrum of NLS1
 \citep[e.g. ][]{gallo04, grupe04c, tanaka05}.
Its presence is also indicated by narrow absorption lines (which
appear to be saturated but do not reach zero intensity) in
UV spectra of Broad Absorption Line (BAL) quasars
\citep[e.g. ][]{barlow97, hamann98, wills99}.
However, the geometry and physics of partial coverers
are still not well understood. One possible geometry
consists of thick blobs of gas, partially covering
parts of the accretion disk \citep[e.g. ][]{guilbert88}.
In the case of Mkn 335, the clouds must cover only the inner
parts of the disk, since we find that the UV emission
is not highly variable between the \xmm\ observation in 2000
and the \swift\ observations in 2007, while the X-rays vary
dramatically\footnote{We note that historic light curves
from IUE and HST did show that Mkn 335 has been
variable in the UV between 1978 and 1985
\citep{dunn06, edelson90} by a factor of 2.
The UBVRI photometry of Mkn 335 as reported by \citet{doroshenko05}
(see also \citet{czerny07}) also suggests that the AGN is intrinsically
highly variable.
However, because of the lack of simultaneous X-ray observations during
these time periods
we do not know if the UV
variability was caused by changes in the flux of the central engine or was
caused by
absorption. }.
Note that the fits to the May 17 and May 25 Swift spectra suggest a change in
the partial covering absorber column densities by a factor of about 2. 
This timescale is consistent with e.g. the absorber toy model suggested by 
\citet{abrassart00}, where thick clouds at 10-100 Schwarzschild radii partially obcure 
the central region and causing the X-ray variability.

Alternatively, a partial covering situation may
arise if our line-of-sight passes through an accretion-disk driven
wind which is launched at intermediate disk radii 
\citep[e.g. ][]{elvis00,proga07}.
If such a wind varies with time 
and/or is inhomogeneous, different parts of the central source
would be covered at different times. 
In both partial covering geometries,
the physics is still uncertain.  In the case
of dense blobs: how are they confined and what
is their origin \citep[e.g. ][]{kuncic97}. 
In case of disk-driven winds: what is the driver
of these massive outflows \citep[e.g. ][]{proga07}.

The high column density
we need in our SWIFT spectral fits is similar to those
frequently observed in BAL quasars \citep[e.g. ][]{green96, gallagher02, grupe03}.
In this context, it is interesting to note that similarities
between NLS1 galaxies and BAL quasars have been pointed out repeatedly
\citep[e.g. ][]{mathur00, bg00, bor02}. 
In one specific case, that of the X-ray transient NLS1 galaxy
WPVS 007, the onset of heavy X-ray absorption
\citep{grupe07b} is indeed accompanied by the onset of
UV BALs \citep{leighly07}.
When correcting for the effects of intrinsic absorption we found that the
X-ray flux of Mkn 335 originating
 from the central engine has
been very similar in the rest-frame 0.2-2.0 keV band between the ROSAT
observations and the most recent \swift\
observations. Using these fluxes, the intrinsic variability is only a
factor of about
 4-6, which is quite normal for an
AGN, in particular for a NLS1. The change in intrinsic flux between the first and 
second Swift observations is about a factor of 3 within a week. If a partial covering absorber is 
the correct model this flux change implies that the soft X-ray scattering region can only be 
a few light days in diameter which is consistent with the \citet{abrassart00} toy model. 
The deep low-state of Mkn 335 discovered with \swift\
provides us with a rare chance
to scrutinize the properties of X-ray low-state AGN in general.
Mkn 335 is unique with respect to being 
relatively bright during its low-state. Therefore, follow-up observations of Mkn 335 in its current
low-state are highly encouraged. We will continue our monitoring with \swift\ in order to 
find the timescales on which the AGN switches from a low to high state, but also deep 
 \xmm\ observations, optical spectrocsopy and spectropolarimetry are needed to clarify the 
 nature of the current low-state.

\acknowledgments
We want to thank Paul O'Brien and our referee
for useful comments on the manuscript.
We also want to thank the \swift\ PI Neil Gehrels for approving our ToO request
and the \swift\ Science Planners Mike Stroh and 
 Sally Hunsberger for fitting the observations into
the \swift schedule.  
We acknowledge the use of public data from the \xmm\ and Swift data archives.
This research has made use of the NASA/IPAC Extragalactic
Database (NED) which is operated by the Jet Propulsion Laboratory,
Caltech, under contract with the National Aeronautics and Space
Administration. Swift is supported at PSU by NASA contract NAS5-00136.
This research was supported by NASA contract NNX07AH67G (D.G.).

\clearpage

\begin{deluxetable}{lcccr}
\tabletypesize{\tiny}
\tablecaption{\xmm\ \& \swift\  observations of Mkn 335
\label{obs_log}}
\tablewidth{0pt}
\tablehead{
\colhead{Mission} & \colhead{T-start\tablenotemark{1}} & 
\colhead{T-stop\tablenotemark{1}} &
\colhead{$\rm T_{exp}$\tablenotemark{2}} 
} 
\startdata
XMM 2000 & 12-25 17:18 & 12-26 02:06 & 36910 \\
XMM 2006 & 01-03 19:10 & 01-05 08:03 & 133251 \\
\swift\ 001/2007 & 05-17 00:32 & 05-17 05:37 &  4860 \\
\swift\ 002/2007 & 05-25 00:01 & 05-25 19:23 & 8084 \\
\swift\ 003/2007 & 06-28 00:01 & 06-28 14:37 & 2932 \\
\swift\ 004/2007 & 06-30 00:13 & 06-30 14.52 & 2837 \\
\swift\ 005/2007 & 07-02 14:47 & 07-02 21:18 & 2979 
\enddata

\tablenotetext{1}{Start and End times are given in UT}
\tablenotetext{2}{Observing time given in s}
\end{deluxetable}

\begin{deluxetable}{ccccccccccccc}
\tabletypesize{\tiny}
\tablecaption{Spectral analysis of the \xmm\ and \swift\ X-ray data. 
\label{xray_res}}
\tablewidth{0pt}
\tablehead{
\colhead{Observation} & 
\colhead{Model\tablenotemark{1}} &
\colhead{$\alpha_{\rm X,soft}$} 
& \colhead{$E_{\rm break}$\tablenotemark{2}}
& \colhead{$\alpha_{\rm X,hard}$}
& \colhead{$N_{\rm H, pcf}$\tablenotemark{3}}
& \colhead{$\rm F_{cover}$\tablenotemark{4}}
& \colhead{log $F_{\rm X,gal}$\tablenotemark{5}}
& \colhead{log $F_{\rm X,all}$\tablenotemark{6}}
& \colhead{$\chi^2/\nu$}
} 
\startdata
XMM 2000 & (a) & 1.87\plm0.01 & 1.76$^{+0.09}_{-0.08}$ & 1.18\plm0.04 & --- &  --- & --13.10 & & 513/419 \\
         & (b) & 1.87\plm0.01 & --- & --- & 7.7$^{+1.1}_{-0.9}$ & 0.58\plm0.02 & --13.10 & --12.72  & 561/419 \\
XMM 2006 & (a) & 1.73\plm0.01 & 1.82\plm0.02 & 1.08\plm0.01 & --- & --- & --13.17 & & 2420/1327 \\
	 & (c) & 1.74\plm0.01 & 1.63\plm0.03 & 1.25\plm0.02 &  55.1$^{+10.0}_{-8.2}$ &
	 0.45$^{+0.04}_{-0.03}$ & --13.17 & --12.91 & 1973/1325 \\
         & (d) & 1.74\plm0.01 & 1.62\plm0.03 & 1.25\plm0.02 & 51.0$^{+10.2}_{-8.0}$ & 0.43$^{+0.04}_{-0.03}$ & --13.17 &
	 --12.93 & 2611/1910\tablenotemark{8} \\
\swift\ 001 & (c) & 2.05$^{+0.24}_{-0.20}$ & --- & --- & 31.1$^{+17.0}_{-13.6}$ & 0.98$^{+0.02}_{-0.03}$ 
	    & --14.63 & & 9/9 \\
         & (e) & 1.74\plm0.01 & 1.62\plm0.03 & 1.25\plm0.02 & 38.2$^{+19.2}_{-19.6}$ & 0.95$^{+0.04}_{-0.03}$ & --14.75 &
	 --13.48 & 2611/1910\tablenotemark{8} \\
            & (f) & 1.79\plm0.10 & --- & --- & 20.0$^{+8.9}_{-5.4}$ & 0.93\tablenotemark{7} & --14.71 & --13.55 & 
	    87/85\tablenotemark{9} \\
         & (g) & 1.81\plm0.10 & --- & --- & 20.5$^{+8.6}_{-5.4}$ & 0.94$^{+0.01}_{-0.02}$ & --14.70 & --13.51 & 87/86\tablenotemark{9} \\
         & (h) & 1.81\plm0.10 & --- & --- & 12.8$^{+2.6}_{-2.2}$ & 0.91$^{+0.03}_{-0.06}$ & --14.72 & --13.69 & 95/86\tablenotemark{9} \\
\swift\ 002 & (c) & 1.78\plm0.14 & --- & --- & 10.4$^{+2.6}_{-2.2}$ & 0.93$^{+0.02}_{-0.03}$ & --14.25 & --13.10 & 52/48  \\
         & (e) & 1.74\plm0.01 & 1.62\plm0.03 & 1.25\plm0.02 & 10.4$^{+3.6}_{-2.7}$ & 0.88$^{+0.02}_{-0.03}$ & --14.27 &
	 --13.37 & 2611/1910\tablenotemark{8} \\
          & (f) & 1.79\plm0.12 & --- & --- & 10.4$^{+2.6}_{-2.1}$ & 0.93\plm0.02 & --14.25 & --13.10 & 87/85\tablenotemark{9} \\
         & (g) & 1.81\plm0.10 & --- & --- & 10.9$^{+2.4}_{-2.0}$ & 0.94$^{+0.01}_{-0.02}$ & --14.25 & --13.06 & 87/86\tablenotemark{9} \\
         & (h) & 1.81\plm0.10 & --- & --- & 12.8$^{+2.6}_{-2.2}$ & 0.94$^{+0.01}_{-0.02}$ & --14.25 & --13.02 & 95/86\tablenotemark{9} \\
\swift\ 003-005\tablenotemark{10} & (c) & 1.75\plm0.17 & --- & --- & 16.6$^{+6.4}_{-4.7}$ & 0.93$^{+0.02}_{-0.04}$ & 
	    --14.49 & --13.32 & 23/26 \\
         & (e) & 1.74\plm0.01 & 1.62\plm0.03 & 1.25\plm0.02 & 17.5$^{+9.8}_{-6.6}$ & 0.89$^{+0.04}_{-0.07}$ & --14.50 &
	 --13.56 & 2611/1910\tablenotemark{8} \\
            & (f) & 1.79\plm0.10 & --- & --- & 16.4$^{+6.1}_{-4.6}$ & 0.94$^{+0.02}_{-0.03}$ & --14.48 & --13.28 &
	    87/85\tablenotemark{9} \\
         & (g) & 1.81\plm0.10 & --- & --- & 15.8$^{+4.0}_{-3.2}$ & 0.94$^{+0.01}_{-0.02}$ & --14.47 & --13.28 & 87/86\tablenotemark{9} \\
         & (h) & 1.81\plm0.10 & --- & --- & 12.8$^{+2.6}_{-2.2}$ & 0.92$^{+0.02}_{-0.03}$ & --14.48 & --13.36 & 95/86\tablenotemark{9} 
\enddata

\tablenotetext{1}{Spectral models used are: (a) absorbed power law, 
(a) absorbed broken power law, 
(b) partial covering absorbed with a single power law,
(c) partial covering absorber and broken power law, (d) same as (c) 
but simultaneous fits to the 2006 \xmm\ and all \swift\ spectra with the broken
power law parameters tied and the partial covering absorber parameters left free to vary,
(e) same as (b) but \swift spectra fit simultaneously in XSPEC with the X-ray spectral index tied and the 
partial covering absorber parameters left free to vary, (f) same as (e) but the covering fraction $f_c$ of all three \swift\ spectra tied, 
and (g) same as (e) but $N_{\rm H}$ tied and $f_c$ left free.
For all models the absorption column density was fixed to the Galactic value
\citep[3.96$\times 10^{20}$ cm$^{-2}$][]{dic90}.  }
\tablenotetext{2}{The break energy $E_{\rm break}$ is given in units of keV.}
\tablenotetext{3}{Absorption column density of the redshifted
partial covering absorber $N_{\rm H, pcf}$ in units of $10^{22}$ cm$^{-2}$}
\tablenotetext{4}{Covering fraction $\rm F_{cover}$}
\tablenotetext{5}{Rest-frame 0.2-2.0 X-ray flux log $F_{\rm 0.2-2.0 keV}$ corrected for Galactic absorption
 given in units of W m$^{-2}$}
\tablenotetext{6}{Rest-frame 0.2-2.0 X-ray flux log $F_{\rm 0.2-2.0 keV}$ corrected for Galactic and intrinsic absorption
 given in units of W m$^{-2}$}
\tablenotetext{7}{Leaving  covering absorber fraction as a free only gives an unconstrained results. 
We therefore fixed the absorption covering fraction to 0.93 which was found in the other  \swift\ data.}
\tablenotetext{8}{Simultaneous fit to all \swift\ data}
\tablenotetext{9}{Simultaneous fit to all \swift\ data}
\tablenotetext{10}{Coadded data from segments 003 to 005}
\end{deluxetable}

\clearpage



\end{document}